\documentstyle[prl,aps,preprint,epsfig]{revtex}

\begin{document}

\addtolength {\oddsidemargin} {-0.0cm}
\addtolength {\topmargin} {1cm}
\setlength{\parindent}{0pc}
\draft

\title{Defibrillation via the Elimination of Spiral Turbulence in a 
Model for Ventricular Fibrillation}

\author{ Sitabhra Sinha${}^{1, 2}$, Ashwin Pande${}^1$
and Rahul Pandit${}^{1, 2}$}

\address{${}^1$ Centre for Condensed Matter Theory,
Department of Physics, Indian Institute of Science,
Bangalore 560 012, India\\
${}^2$ Condensed Matter Theory Unit, Jawaharlal Nehru Centre for
Advanced Scientific Research, Bangalore 560 064, India}

\maketitle

\widetext

\begin{abstract}

Ventricular fibrillation (VF), the major reason behind sudden 
cardiac death, is turbulent cardiac electrical activity in which
rapid, irregular disturbances in the spatiotemporal electrical 
activation of the heart makes it incapable of any concerted pumping 
action. Methods of controlling VF include electrical 
defibrillation as well as injected medication. Both methods yield 
results that are subject to chance. Electrical defibrillation, though 
widely used, involves subjecting the whole heart to massive, and 
often counterproductive, electrical shocks. We propose a defibrillation 
method that uses a very-low-amplitude shock (of order mV) applied 
for a brief duration (of order 100 ms) and over a coarse mesh of 
lines on our model ventricle. Our proposal is based on a detailed 
numerical study of a model for ventricular fibrillation\cite{Pan93}.
\end{abstract}


\narrowtext

Ventricular fibrillation (VF), the leading cause of sudden cardiac 
death, is responsible for about one out of every six deaths in the
USA\cite{Win98,website}. In the absence of any attempt at defibrillation, 
VF leads to death in a few minutes. The importance of this problem can hardly 
be overemphasized, so it is not surprising that many studies of VF have 
been carried out in various mammalian hearts\cite{Win98} and 
mathematical models\cite{models}. Recent experiments\cite{Jal98,Gra98} 
have shown that, though VF is turbulent cardiac activity, it is 
associated with the formation, and subsequent breakup, of 
electrophysiological structures that emit spiral waves; these are 
referred to as rotors or simply spirals. Such structures and 
their breakup have also been obtained in a set of partial differential 
equations of the Fitzhugh-Nagumo type; this set has been 
proposed as a model for VF\cite{Pan93}.  We show that the spiral 
turbulence associated with VF in this model arises because of 
{\em spatiotemporal chaos}; this is similar to the spatiotemporal 
chaos in a related model\cite{Hil95} for the catalysis of CO on 
Pt(110) which we have studied elsewhere\cite{Pan99,Pan00}. 
Our defibrillation scheme for the VF model of Ref.\cite{Pan93} 
relies on the understanding we have developed for spatiotemporally 
chaotic states in the CO catalysis model.

Electrical defibrillation entails the application of electrical jolts 
to the fibrillating heart to make it start beating normally again.
Such defibrillation works only about two-thirds of the time and often 
damages the heart in the process of reviving it\cite{Poo90}. In 
external defibrillation, electrical shocks ($\simeq 5$ kV) are 
applied across the patient's chest; they depolarise all heart cells 
simultaneously and essentially reset the pacemaking nodes of the 
heart\cite{Eis86}. Slightly lower voltages ($\simeq 600$ V) 
suffice in {\em open-heart} conditions, i.e., when the shock is 
applied directly on the heart's surface\cite{Pil70}. In our 
study of the VF model\cite{Pan93} we achieve 
defibrillation by using 
{\em very-low-amplitude} pulses (of order mV) applied for
a {\em brief} duration (of order 100 ms) and over a coarse mesh 
of lines on our model ventricle. Thus, if our defibrillation scheme can
be realised in an internal defibrillator, it would constitute a 
significant advance in these 
devices.  

Before proceeding to our quantitative results it is useful to 
define the model for ventricular fibrillation\cite{Pan93}.
For simplicity we use the model for isotropic cardiac tissue; in this 
case the equations governing the excitability $e\/$ and recovery 
$g\/$ variables are
\begin{equation}
\begin{array}{lll}
{\partial e}/{\partial t} & = & {\nabla}^2 e - f(e) - g,\\
{\partial g}/{\partial t} & = & {\epsilon}(e,g) (ke - g).
\end{array}
\end{equation}
The function $f(e)\/$, which specifies fast processes (e.g., 
the initiation of the action potential) is piecewise linear:
$f(e)= C_1 e$, for $e<e_1$, $f(e) = -C_2 e + a$, 
for $e_1 \leq e \leq e_2$, and $f(e) = C_3 (e - 1)$, for $e > e_2$. 
We use the physically appropriate parameter values given 
in Ref.\cite{Pan93}, namely, $e_1 = 0.0026$, $e_2 = 0.837$, $C_1 = 20$, 
$C_2 = 3$, $C_3 = 15$, $a = 0.06$ and $k = 3$. The function 
$\epsilon (e,g)$, which determines the dynamics of the
recovery variable, is $\epsilon (e,g) = \epsilon_1$ for 
$e < e_2$, $\epsilon (e,g) = \epsilon_2$ for $e > e_2$, and
$\epsilon (e,g) = \epsilon_3$ for $e < e_1$ and $g < g_1$ with 
$g_1 = 1.8\/$, $\epsilon_1 = 1/75\/$, $\epsilon_2 = 1.0\/$, and 
$\epsilon_3=0.3$ (which lies in the range $2 > {\epsilon}_3 > 0.1$
suggested in Ref.\cite{Pan93}). 
We solve model (1) by using a forward-Euler integration scheme. We 
discretise our system on a grid of points in space with spacing 
$\delta x = 0.5\/$ dimensionless units and use the standard five-
and seven-point difference stencils\cite{numrec} for the Laplacians
in spatial dimensions $d = 2$ and 3, respectively. Our spatial 
grid consists of a square lattice with $L \times L\/$ points
or cubic lattices with $L \times L \times L_z\/$ points;
in our studies we have used values of $L\/$ ranging from $128\/$ 
to $512\/$ and $2 \leq L_z \leq 16$. 
Our time step is $\delta t = 0.022$ dimensionless units. 
As in Ref.\cite{Pan93}, we define dimensioned time $T$ 
to be $5$ ms times dimensionless time and $1$ spatial unit 
to be $1$ mm, such that the period and wavelength of a spiral wave 
are approximately 120ms and 32.5 mm, respectively. The dimensioned
value of the conductivity constant in model (1) is 2 cm$^2$/s \cite{Pan93}.
Our initial condition is a broken plane wave which we allow to evolve 
into a state displaying spiral turbulence; we control this
eventually by our defibrillation scheme. On the edges of our 
simulation region we use no-flux (Neumann) boundary conditions since the 
ventricles are electrically insulated from the atria. We impose 
these boundary conditions numerically by adding an 
extra layer of grid points on 
each side of our simulation grid and requiring the
values of $e\/$ be equal pointwise 
to their values on the points one layer within the boundary.

Before constructing an efficient defibrillation scheme for VF
in model (1) it is important to appreciate the following
points:
(a) Ventricular fibrillation in this model arises because the system
evolves to a state in which large spirals break down \cite{Pan93,Pan99}.
(b) This state is a long-lived transient whose lifetime $\tau_L$ increases
rapidly with $L$, the linear size of our system
(e.g., for $d = 2$, $\tau_L \simeq 850$ ms for $L = 100$ 
whereas $\tau_L \simeq  3200$ ms for $L = 128$). 
This property of model (1) is in qualitative accord with
the experimental finding that the hearts of small mammals are less prone
to fibrillation than those of large mammals \cite{Win87,Kim97}. For
time $t \gg \tau_L$, a quiescent state with $e = g = 0$ is obtained. (c) In 
systems with $L \gtrsim 128$, $\tau_L$ is sufficiently 
long that a nonequilibrium
statistical steady state is established. This state displays 
spatiotemporal chaos.
For example, we find that there are several positive Lyapunov exponents
$\lambda_i$ (averages for $\lambda_i$ are performed for $\tau_0 < t < 
\tau_L$, where $\tau_0$ is the time of decay for initial 
transients\cite{lyap1}); the number of positive $\lambda_i$ increases with 
$L$ (e.g., the Kaplan-Yorke dimension  $D_{KY}$ \cite{lyap2} 
increases from 7 to $\stackrel{>}{\sim}$ 35 as $L$ increases 
from 128 to 256).
(d) Model (1) is akin to a model for the catalysis of CO on Pt(110)
in so far as both show spiral breakup and spatiotemporal 
chaos \cite{Pan93,Pan99}. Our recent studies of the 
CO catalysis model \cite{Pan00}
have shown that Neumann or
no-flux boundary conditions tend to absorb spiral defects and, indeed, the
spirals do not last for appreciable periods of time on small systems.

Given that the long transient which leads to VF
in model (1) is spatiotemporally chaotic, we might guess that the fields
$e\/$ and $g\/$ have to be controlled globally to achieve defibrillation.
In fact, some earlier studies \cite{Osi99} of spiral breakup in models 
for ventricular fibrillation have used global control.
Here we show that a judicious choice of control points (on a mesh specified
below) leads to an efficient defibrillation scheme for model (1). 
For $d = 2$, we divide our simulation domain 
(of size $L \times L$) into $K^2$
smaller blocks and choose the
mesh size such that it effectively suppresses the formation
of spirals. For $d = 3$, we use the same control mesh {\em but only
on one of the square faces} of 
the $L \times L \times L_z\/$ simulation box.
In our defibrillation scheme we apply a pulse to the $e\/$ field 
on a 
mesh composed of lines of width  $3 \delta x$. A network of such
lines is used to divide 
the region of simulation into square blocks whose length in each 
direction is fixed at a constant value $L/K$ for the duration of control. 
(The blocks adjacent to the boundaries can turn be out to be rectangular).
The essential point here is that, if a pulse is applied to
the $e\/$ field at all points along the mesh boundaries for a time $\tau_c$,
then it effectively simulates Neumann boundary conditions (for the
block bounded by the mesh) in so far as it absorbs spirals formed
inside this block (just like Neumann boundary conditions absorb
spiral defects in the CO catalysis model \cite{Pan00}).
Note that $\tau_c$ is not large at all since the
individual blocks into which the mesh divides our 
system are of a linear size $L/K$ which is so small that it does not
sustain long, spatiotemporally chaotic transients. Nor does $K$,
which is related to the mesh density, have to be very large
since the transient lifetime, $\tau_L$, decreases rapidly with
decreasing $L$. We find that, for $d = 2$, $L = 128$, $K = 2$ 
and $\tau_c = 41.2$ ms is required
for defibrillation. In Fig. 2  we illustrate such defibrillation 
with $\tau_c = 44$ ms. For $d = 2$, $L = 512$, $K = 8$
and $\tau_c = 704$ ms suffices. Finally
we show that a slight modification of our defibrillation scheme 
also works for $d = 3$ (Fig. 3).

The efficiency of our defibrillation scheme is fairly insensitive
to the height of the pulse we apply to the $e\/$ field along our control
mesh so long as this height is above a threshold.
To obtain the value of $e$ in mV units we have scaled the peak
amplitude of a spike in the $e\/$ field (which has an amplitude of
0.9 in dimensionless units) to be 
equal to 110 mV. The latter
is a representative average
value of the peak voltage of
an electrical wave in the heart \cite{Win87}. With this voltage scaling, 
dimensioned excitability is computed as $110/0.9 \simeq 122.22$ mV times 
dimensionless excitability. We find, e.g., that, for $L = 128\/$, the 
smallest pulse which yields defibrillation is 57.3 mV/ms for the 
parameter values we use; however, we have checked
that even stronger pulses (e.g., 278.3 mV/ms) 
also lead to defibrillation. We use a capacitance density 
of 1 $\mu$F/cm$^2$ \cite{Bee77}, which then yields a current 
density of 57.3 $\mu$A/cm$^2$. Note that this
threshold is of the order of the smallest potential 
($\simeq 22.18\/$ $\mu$A/cm$^2$)
required to trigger an action-potential spike in model (1) 
with $g = 0\/$; the nullclines for model (1), 
in the absence of the Laplacian, are such
that this smallest potential increases with increasing $g\/$. This is
physically consistent with the increase in the refractory nature of
the heart tissue with increasing $g\/$.

We have checked that (a) small, local deformations of our control mesh
or (b) the angle of the mesh axes with the boundaries do not affect the
efficiency of our defibrillation scheme. Furthermore the application of 
our control pulse on the control mesh does not lead to an instability 
of the 
quiescent state; thus it cannot inadvertently promote VF. We have checked
specifically that, with $e = 0$, $g = 0$ 
as the 
initial condition at all spatial 
points, a wave of activation travels across the system 
when the control pulse is initiated; this travels quickly
($\simeq$ 200 ms for $L$ = 128) to the boundary 
where it is absorbed and quiescence 
is restored. Our defibrillation method also works if 
$g$ is stimulated instead of $e$.
This can be implemented by pharmaceutical means in an actual heart.

Our two-dimensional defibrillation scheme above applies without
any change to {\it thin} slices of cardiac tissue.
However, it is important to investigate whether it can
be extended to three dimensions which is clearly required for real
ventricles. A naive extension of our mesh into a cubic array of
sheets will, of course, succeed in achieving defibrillation. However,
such an array of control sheets cannot be easily implanted
in a ventricle. We have tried to see, therefore, if we can control
the turbulence in a three-dimensional version of model (1) on a
$L \times L \times L_z\/$ domain {\it but with the control mesh
present only on one} $L \times L$ {\it face}. For the open faces
we use open boundary conditions and for the other faces
we use no-flux Neumann boundary conditions.
Our defibrillation scheme works if $L_z \leq 4$ (we
have checked explicitly for $L = 220$) but not for $L_z > 4$.
A slight modification of this scheme is
effective even for $L_z > 4$: Instead of applying a pulse for a
duration $\tau_c$, we apply a sequence of $n$ pulses separated by
a time $\tau_{ip}$ and each of duration $\tau_w$. We find that, if
$L = 256$ and $L_z = 8$, defibrillation occurs in $\simeq 2002$ ms
with $\tau_{ip} = 22$ ms, $\tau_w = 0.11$ ms, $n = 30$ and a control
pulse amplitude of 57.3 $\mu$A/cm$^2$; if $L = 128$
and $L_z = 16$, defibrillation occurs in $\simeq 1760$ ms with
$\tau_{ip} = 22$ ms, $\tau_w = 0.11$ ms, $n = 15$ and
a control pulse amplitude of 57.3 $\mu$A/cm$^2$. We further
find that optimal defibrillation is obtained in our model
if $\tau_{ip}$
is close to the absolute refractory period for model (1)
without the Laplacian term. The efficacy of our control scheme
in the three-dimensional case can be understood heuristically
as follows: Control via a steady pulse does not work for $L_z > 4$
since the propagation of this pulse in the $z$ direction
(normal to our control mesh) is blocked once the medium in the
interior of our simulation domain becomes refractory. However, if we
use a sequence of short pulses separated by a time $\tau_{ip}$, then,
provided $\tau_{ip}$ is long enough for the medium to recover its
excitability, the control-pulse waves can propagate in the $z$ 
direction and lead to successful defibrillation. 

Typical electrical defibrillation schemes
use much higher voltages than in our study. Recent
studies \cite{Osi99,Rap99} have explored low-amplitude defibrillation
methods in model systems. They constitute an advance over
conventional methods, but lack some of the appealing features
of our defibrillation scheme. For example, the scheme of 
Ref.\cite{Osi99} works only when the slow variable (the analog
of our $g$) is controlled; though this can be done, in principle,
by pharmaceutical means, it is clearly less direct than control
via electrical means. Reference \cite{Rap99}
uses electrical defibrillation, but has been demonstrated to prevent only
one spiral from breaking up, as opposed to the suppression of a
spatiotemporally chaotic state with broken spirals by our defibrillation
scheme. We have checked explicitly that, for the spatiotemporally chaotic 
state of model (1), a straightforward implementation of the defibrillation
scheme of Ref.\cite{Rap99} is ineffective. [This scheme applies pulses
to the fast variable ($e$ in model (1)) on a two-dimensional, discrete
lattice of points.] The control current density in Ref. \cite{Rap99}
is comparable to our $\simeq 57.3 \mu$A/cm$^2$, which is much lower than
139$\mu$A/cm$^2$, the maximum value of the ionic current
during depolarization in the Beeler-Reuter model.

We have checked that our defibrillation scheme is not sensitively
model dependent. For example, we have used the same scheme to eliminate 
spatiotemporal chaos associated with spiral breakup in the model for
the catalysis of CO on Pt(110) mentioned above\cite{Hil95,Pan99}. 
We have found recently that our defibrillation scheme also works
for the biologically realistic Beeler-Reuter model \cite{Bee77,Sin00}
for VF.

In conclusion, then, we have developed an efficient
method for defibrillation by the elimination of spiral
turbulence in model (1).
Our method has the
attractive feature that it uses very-low-amplitude pulses,
applied only for a short duration on a coarse
control mesh of lines. And, to the best of our knowledge, our
study is the only one which shows how to attain defibrillation 
by the control of spatiotemporal chaos and spiral turbulence 
in a model for VF in both two and three dimensions. 
We hope our work will stimulate
experimental tests of the efficacy of our defibrillation method.

\vspace{0.5cm}
We thank CSIR (India) for support,
SERC (IISc, Bangalore) for 
computational facilities, and A. Pumir and N.I. Subramanya for 
discussions.

\pagebreak

{\large \bf Figure Legends}

\vspace{1cm}

{\bf Figure 1.}
Pseudograyscale plots showing the evolution of spatiotemporal
chaos through spiral breakup in model (1) for physical times $T$ between
880 ms and 1430 ms. The panels on the left show the excitability $e\/$
and those on the right the recovery $g\/$ for all points ($x$,$y$) on a
two-dimensional $L \times L$ spatial grid with $L = 128$. The initial
condition used is described in the text.

\vspace{1cm}
{\bf Figure 2.}
Pseudograyscale plots of the $e\/$ field in $d$ = 2 for
$L$ = 128 (top) and $L$ = 512 (bottom) illustrating defibrillation
by our control of spiral breakup in model (1).
The control mesh divides the domain into four
equal squares for $L$ = 128 and, for $L$ = 512, into 49 equal squares,
28 equal rectangles along the edges and 4 small equal squares on
the corners. For $L$ = 128 we apply 57.3 $\mu$A/cm$^2$ from $T$ = 891 ms to
$T$ = 935 ms and by $T$ $\stackrel{>}{\sim}$ 1500 ms spatiotemporal
chaos is all but eliminated ($|e|, |g| \leq 10^{-13}$ at all grid points).
For $L$ = 512, we apply 250 $\mu$A/cm$^2$ from $T$ = 55 ms
to $T$ = 759 ms to the
$e\/$ field. By $T$ = 1650 ms, spatiotemporal chaos is all
but eliminated ($|e|, |g| \leq 10^{-7}$ at all grid points)
(see
http://theory1.physics.iisc.ernet.in/rahul/images/heart.mpg
for an animated figure).

\vspace{1cm}

{\bf Figure 3.}
Isosurface ($e = 0.6$) plots of initial states (left panels)
and pseudograyscale plots of the $e$ field on the top square face
(right panels) illustrating defibrillation by the control
of spiral breakup in model (1) for $d = 3$: (top) $L = 128$ and
$L_z = 16$, with the control mesh dividing the top face into 4
equal squares; and (bottom) $L = 256$ and $L_z = 8$, with the control
mesh dividing the top face into 64 equal squares. In both cases we
apply pulses on the control mesh ({\it top face only}) of 57.3 $\mu$A/cm$^2$
with $\tau_{ip} = 22$ ms and $\tau_w = 0.11$ ms
(see text). In the former
case, with $n = 15$, spatiotemporal chaos is all but eliminated by
$T$ = 1760 ms ($|e|, |g| \leq 10^{-4}$ at all grid points); in the
latter, with $n = 30$, it is all but eliminated by $T$ = 2002 ms
($|e|, |g| \leq 10^{-3}$ at all grid points).


\begin{thebibliography}{10}
\bibitem{lyap1} We use standard algorithms\cite{Pan99,Chu89} for 
calculating the first $N\/$ Lyapunov exponents with $N = 35\/$ in 
most of our studies.
\bibitem{lyap2} We calculate the Kaplan-Yorke 
dimension\cite{Chu89} as follows:
Label the calculated Lyapunov exponents $\lambda_i\/$ in decreasing 
order such that $\lambda_1 \geq \lambda_2 \geq \lambda_3 \cdots \geq 
\lambda_N\/$. Now define $D\/$ as that index at which
$\sum_{i=1}^{D} \lambda_i \geq 0$ and $\sum_{i=1}^{D+1} \lambda_i < 
0$. Then the Kaplan-Yorke dimension $D_{KY}\/$ is
$D_{KY}= D + {(\sum_{i=1}^{D} \lambda_i)}/{|\lambda_{D+1}|}$.
\end{thebibliography}
\end{document}